\documentclass[prb,twocolumn,showpacs,amsmath,amssymb]{revtex4}
\usepackage{times}
\usepackage{mathptmx}
\usepackage{dcolumn}
\usepackage{epsf}
\usepackage{graphicx}
\epsfverbosetrue

\def\rmd     {\hbox{\scriptsize d}}
\def\rmdMP     {\hbox{\scriptsize d:MP}}
\def\rmC     {\hbox{\scriptsize C}}
\def\rmT     {\hbox{\scriptsize T}}
\def\rmRx    {\hbox{\scriptsize :Rx}}
\def\rmId    {\hbox{\scriptsize :Id}}
\def\Or      {\hbox{O}\!}

\def\VPs     {$\hbox{V}_{\hbox{\scriptsize P}}$ }

\def\PIq	{$\hbox{P}_{\hbox{\scriptsize In}}^{\hbox{\scriptsize +2}}$}

\def\VPq	{$\hbox{V}_{\hbox{\scriptsize P}}^{\hbox{\scriptsize +1}}$}
\def\IiPq	{$\hbox{In}_{\hbox{\scriptsize i(P)}}^{\hbox{\scriptsize +3}}$}
\def\IiPn	{$\hbox{In}_{\hbox{\scriptsize i(P)}}^{\hbox{\scriptsize +0}}$}
\def\PiPq	{$\hbox{P}_{\hbox{\scriptsize i(P)}}^{\hbox{\scriptsize +3}}$}
\def\IPq	{$\hbox{In}_{\hbox{\scriptsize P}}^{\hbox{\scriptsize -2}}$}
\def\VIq	{$\hbox{V}_{\hbox{\scriptsize In}}^{\hbox{\scriptsize -3}}$}

\def\SPq	{$\hbox{S}_{\hbox{\scriptsize P}}^{\hbox{\scriptsize +1}}$}
\def\SiIq	{$\hbox{Si}_{\hbox{\scriptsize In}}^{\hbox{\scriptsize +1}}$}
\def\ZnIq	{$\hbox{Zn}_{\hbox{\scriptsize In}}^{\hbox{\scriptsize -1}}$}
\def\ZniPq	{$\hbox{Zn}_{\hbox{\scriptsize i(P)}}^{\hbox{\scriptsize +2}}$}
\def\SiPq	{$\hbox{Si}_{\hbox{\scriptsize P}}^{\hbox{\scriptsize -1}}$}

\def\PIqs	{$\hbox{P}_{\hbox{\scriptsize In}}^{\hbox{\scriptsize +2}}$ }
\def\PIns	{$\hbox{P}_{\hbox{\scriptsize In}}^{\hbox{\scriptsize +0}}$ }

\def\IiPqs	{$\hbox{In}_{\hbox{\scriptsize i(P)}}^{\hbox{\scriptsize +3}}$ }

\def\IPqs	{$\hbox{In}_{\hbox{\scriptsize P}}^{\hbox{\scriptsize -2}}$ }
\def\VIqs	{$\hbox{V}_{\hbox{\scriptsize In}}^{\hbox{\scriptsize -3}}$ }
\def\VIns	{$\hbox{V}_{\hbox{\scriptsize In}}^{\hbox{\scriptsize +0}}$ }

\def\SiIqs	{$\hbox{Si}_{\hbox{\scriptsize In}}^{\hbox{\scriptsize +1}}$ }
\def\ZnIqs	{$\hbox{Zn}_{\hbox{\scriptsize In}}^{\hbox{\scriptsize -1}}$ }

\def\ZnI      {$\hbox{Zn}_{\hbox{\scriptsize In}}$}
\def\ZnIs     {$\hbox{Zn}_{\hbox{\scriptsize In}}$ }

\def\ZniP      {$\hbox{Zn}_{\hbox{\scriptsize i(P)}}$}

\def\SiI      {$\hbox{Si}_{\hbox{\scriptsize In}}$}
\def\SiIs     {$\hbox{Si}_{\hbox{\scriptsize In}}$ }
\def\SiP      {$\hbox{Si}_{\hbox{\scriptsize P}}$}

\def\SPs     {$\hbox{S}_{\hbox{\scriptsize P}}$ }

\def\SPT     {$\hbox{S}_{\hbox{\scriptsize P}}^{\hbox{\scriptsize +/0}}$}
\def\SiIT     {$\hbox{Si}_{\hbox{\scriptsize In}}^{\hbox{\scriptsize +/0}}$}
\def\SiPT     {$\hbox{Si}_{\hbox{\scriptsize P}}^{\hbox{\scriptsize 0/-}}$}
\def\ZnIT     {$\hbox{Zn}_{\hbox{\scriptsize In}}^{\hbox{\scriptsize 0/-}}$}
\def\SPTs     {$\hbox{S}_{\hbox{\scriptsize P}}^{\hbox{\scriptsize +/0}}$ }
\def\SiITs     {$\hbox{Si}_{\hbox{\scriptsize In}}^{\hbox{\scriptsize +/0}}$ }
\def\SiPTs     {$\hbox{Si}_{\hbox{\scriptsize P}}^{\hbox{\scriptsize 0/-}}$ }
\def\ZnITs     {$\hbox{Zn}_{\hbox{\scriptsize In}}^{\hbox{\scriptsize 0/-}}$ }

\def\SPDT     {$\hbox{S}_{\hbox{\scriptsize P}}^{\hbox{\scriptsize +2/+}}$}
\def\SiIDT     {$\hbox{Si}_{\hbox{\scriptsize In}}^{\hbox{\scriptsize +2/+}}$}
\def\SiPDT     {$\hbox{Si}_{\hbox{\scriptsize P}}^{\hbox{\scriptsize -/-2}}$}
\def\ZnIDT     {$\hbox{Zn}_{\hbox{\scriptsize In}}^{\hbox{\scriptsize -/-2}}$}

\begin{document}

\title{Managing the supercell approximation for charged defects in semiconductors: 
finite size scaling, charge correction factors, the bandgap problem and the {\it\bf ab initio} dielectric constant.}

\author{C W M Castleton$^{1,2,\dagger}$, A.H\"oglund$^{3}$ and S Mirbt$^{3}$}

\address{1 Material Physics, Materials and Semiconductor Physics Laboratory, 
Royal Institute of Technology (KTH), Electrum 229, SE-16440 Kista, 
Sweden.} 
\address{ 2 Department of Physical Electronics/Photonics, 
ITM, Mid Sweden University, SE-85170 Sundsvall, Sweden.} 
\address{ 3 Theory of Condensed Matter, Department of Physics, Uppsala University, 
Box 530, SE-75121 Uppsala, Sweden.}

\date{\today}

\begin{abstract}
The errors arising in {\it ab initio} density functional theory studies of semiconductor point defects using the supercell approximation are analyzed. It is demonstrated that a) the leading finite size errors are inverse linear and inverse cubic in the supercell size, and b) finite size scaling over a series of supercells gives reliable isolated charged defect formation energies to around $\pm$0.05 eV.
The scaled results are used to test three correction methods. The Makov-Payne method is insufficient, but combined with the scaling parameters yields an {\it ab initio} dielectric constant of 11.6$\pm$4.1 for InP. $\Gamma$ point corrections for defect level dispersion are completely incorrect, even for
 shallow levels, but re-aligning the total potential in real-space between defect and bulk cells actually corrects the electrostatic defect-defect interaction errors as well. Isolated defect energies to $\pm$0.1 eV are then obtained using a 64 atom supercell, though this does not improve for larger cells.
Finally, finite size scaling of known dopant levels shows how to treat the band gap problem: in $\le$200 atom supercells with no corrections, continuing to consider levels into the theoretical conduction band (extended gap) comes closest to experiment. However, for larger cells or when supercell approximation errors are removed, a scissors scheme stretching the theoretical band gap onto the experimental one is in fact correct.\end{abstract}
\maketitle

\section{Introduction.} 
Understanding the properties of point defects and dopants is of key importance in studying the electrical and optical properties of semiconductors. While various experimental techniques have been developed over the last half century it is only in recent years that they have started to be matched by accurate first principles computational techniques. Developments in computing power have now made  {\it ab initio} Density Functional Theory\cite{DFT} (DFT) one of the most versatile atomic scale tools available for the investigation of defect properties in semiconductors and insulators. The key quantity to calculate is the defect formation energy
\begin{equation}\label{Eform}
E_{\rmd}^{\rmC} = E_{\rmT}^{\rmC}\!\left(\hbox{defect}^q\right) - 
E_{\rmT}^{\rmC}\!\left(\hbox{no defect}\right) + \sum_i \mu_in_i
- q\left(\epsilon_v + \epsilon_F\right)
\end{equation}
\noindent where $E_{\rmT}^{\rmC}\!\left(\hbox{defect}\right)$ and 
$E_{\rmT}^{\rmC}\!\left(\hbox{no defect}\right)$ are the total energy of the 
supercell ``$\rmC$" with and without the defect, (or charge $q$,) calculated using the same values 
of planewave cutoff, k-point grid, etc, to make use of the cancellation 
of errors.  The defect is formed by adding/removing $n_i$ atoms of 
chemical potential $\mu_i$. $\epsilon_F$ is the Fermi level, measured from $\epsilon_v$, the valence band edge (VBE). Almost all properties of a defect can be derived from variations in and differences between formation energies. The method is very powerful, but critical limitations remain, two of the most important being the relatively small number of atoms which can be treated and the effect of the approximations, such as the Local Density and Generalized Gradient Approximations (LDA and GGA) required to solve the DFT itself. These treat quantum many body correlation and exchange effects incompletely, which in the case of semiconductors and insulators results in a roughly 50\% underestimation of the bandgap. This in turn has severe consequences for the calculation of defect transfer levels, the values of $\epsilon_F$ at which the most stable charge state of the defect changes, given by the difference in $E_{\rmd}^{\rmC}$ between the two states. The positions of the transfer levels govern whether the defect will be a single or multiple acceptor or donor, with levels deep inside the gap, or with shallow levels near to the band edges. Since the predicted band gap differs so strongly from the experimental one it is very hard to map the calculated transfer levels onto the experimental gap and hence predict the electrical properties of the material. 

Meanwhile, the small number of atoms involved (100s or 1000s) means that the boundary conditions become very important. One of the most common approaches is to use Periodic Boundary Conditions (PBCs) together with a plane wave basis set\cite{Payne Review}. A supercell containing the defect in question is repeated periodically throughout space. The cell boundary thus looks bulk-like, rather than being a vacuum as with open boundary conditions. However, it also means that the defect interacts with an infinite array of images of itself seen in the PBCs. This alters $E_{\rmd}^{\rmC}$, making it (and most other defect properties) supercell size dependent. The ``true" defect properties are only recovered in the limit of an infinitely large supercell, equivalent to the limit of an isolated defect. This problem is particularly severe in the case of charged defects, where the Madelung energy becomes infinite if the charge is not neutralized using a uniform jellium background\cite{Jellium}. Even with jellium, the calculated formation energies can be wrong by several eV in supercells of the order of 10s or 100s of atoms, and we have previously shown\cite{NeutralV,NeutralIA} that finite size errors on this scale can even arise for neutral defects. Various authors,\cite{MP,OtherKorr,MP-fails-OtherKorr} have attempted to create corrections schemes to estimate and remove these errors, the most widely known being that of Makov and Payne\cite{MP}.  Although these corrections are often used their accuracy has been strongly questioned, with several studies suggesting that they are not reliably enough for regular use\cite{MP-fails-OtherKorr,MP-fails,VdeW+N}. 

We previously\cite{NeutralV,NeutralIA} suggested that the supercell size errors can instead be eliminated by calculating the same defect properties in a series of supercells of different sizes but the same symmetry and then finite size scaling the results to recover those of the infinite supercell.
We found that $E_{\rmd}^{\rmC}$ varies with the supercell size, 
$L$, as
\begin{equation}\label{Scale_eqn}
E_{\rmd}^{\rmC}\!(L) = E_{\rmd}^{\infty} + a_{1}L^{-1} + a_{n}L^{-n}
\end{equation}
\noindent where $a_{1}$, $a_{n}$ and $E_{\rmd}^{\infty}$ are fitting parameters,  
$E_{\rmd}^{\infty}$ being the finite size scaled formation energy 
corresponding to an infinitely large supercell.  The linear term has been discussed many times previously, first by Leslie and Gillian \cite{Jellium}. For neutral defects we found the correct value for $n$ to be 3. This is actually very intuitive: most sources of error should vary with either the supercell size $L$ (the defect-defect image distance) or with the cell volume $L^3$ (proportional to the jellium charge density, the number of atoms, the number of electrons, etc). Terms scaling with the surface area, $6L^2$, seem unlikely to be dominant.

Here, two further sources of error must also be considered. Firstly, since the electrostatic potential in a supercell with PBCs is only defined up to a constant, the zero on the energy scale must be chosen arbitrarily in each calculation. In the case of most pseudopotential codes (including the one we use) this occurs as an implicit average over values appropriate to each atom species in the supercell, weighted by the number of atoms of each species. This means that the numerical value of $\epsilon_F$ entering Eq \ref{Eform} changes with the contents of the cell, leading to an additional finite size error. If the number of defects per supercell is constant then this error decreases with the number of atoms in the cell - essentially with the volume of the cell, $L^3$. Hence this error is completely taken care of in the infinite supercell limit of our finite size scaling scheme. For individual supercells, 
Van de Walle and Neugebauer\cite{VdeW+N} suggest correcting the error by re-aligning the potential in the defect cell to that of the bulk, using its real-space value at some chosen point in a bulk-like region far from the defect. (We here use the point furthest from the defect in the unrelaxed cell.)

Secondly, additional errors come from the dispersion of the defect levels introduced by overlap between the defect state wavefunctions and their PBC images. It has been suggested\cite{Wei} that this artificially raises $E_{\rmd}^{\rmC}$ when k-points other than just the $\Gamma$ point are used. It is suggested\cite{Wei} that $E_{\rmd}^{\rmC}$ should then be shifted by $q\times\left(\epsilon_D^{\Gamma} - \epsilon_D^{ks}\right)$, where $\epsilon_D^{\Gamma}$ and $\epsilon_D^{ks}$ are the values of the Kohn-Sham level corresponding to the defect state calculated in the defect cell at the $\Gamma$ point and averaged over the sampled k-points respectively. The assumption is that the value of the defect level is correct at the $\Gamma$ point, so the difference between that and the k-point averaged value should be removed. It has been shown by H\"oglund et al.\cite{AndreasGaP} that this is completely incorrect for the example of the phosphorus antisite in GaP. By plotting the ``bandstructure" of the defect level in different sized supercells it was shown that the defect level in the smaller cells is more-or-less correct when averaged over the sampled k-points, but much too low at the $\Gamma$ point. The same is also true for the As vacancy on the GaAs(110) surface, for example\cite{Quasiparticles}. Van de Walle and Neugebauer\cite{VdeW+N} instead point out\cite{VdeW+N} that in this respect there is a fundamental difference between deep levels such as these and shallow defect levels. They suggest that the correction should only be applied when evaluating transfer levels for shallow donors and acceptors.

In the current paper we will show in section \ref{Scaling section} that Eq \ref{Scale_eqn} with $n=3$ also holds for charged defects, so that finite size scaling can be used to produce fully finite-size corrected defect formation and other energies, with well defined error bars and uncertainty. To do this we will study 11 example defects in the zinc-blend structured III-V semiconductor InP. These are chosen to include all types of native defects (vacancies, antisites and interstitials) as well as some common dopants at both substitutional and interstitial sites. Each is studied in one charge state only, usually the one that previous studies\cite{NeutralIA,Previous InP} suggest it has over the majority of the bandgap. The specific choices have been made to include all non-zero values from -3 to +3. 

These results will also enable us in section \ref{Corrections section}~to perform the most objective and comprehensive reliability test we are currently aware of on other, computationally cheaper, correction schemes. (Previous tests rely on only one or two - usually rather simple - examples, and do not generally have reliable isolated formation energies to compare with.) We will test the Makov-Payne scheme, potential re-alignment and dispersion corrections for shallow levels. In section \ref{Ab initio epsilon section} we will also derive an ab initio dielectric constant for InP by combining the scaling results with those of the Makov-Payne correction scheme. Finally, in section \ref{Bandgap section} we will use finite size scaling to provide the first clear-cut answer to the problem of mapping LDA or GGA transfer levels onto the experimental bandgap. Computational details are in the next section, and in section \ref{Conclusions} we will conclude.

\section{Computational Details.} 
We use planewave ab initio DFT \cite{DFT} within the local density 
approximation (LDA) together with ultrasoft pseudopotentials
\cite{USPP} (US-PP) using the VASP code \cite{VASP}. Since we expect (at least) a 3 parameter fit we need at least 4 supercells. These must all be of the same symmetry since the errors scale differently for different symmetries. We choose simple cubic supercells containing 8, 64, 216 and 512 atoms. It would be preferable to replace the 8 atom cell by the 1000 atom one, but our computing resources are currently insufficient for k-point converged calculations with 1000 atoms. On the other hand, we previously found that, somewhat surprisingly, the 8 atom supercell is good enough in most cases: formation energies in this cell usually lie very close to the scaling curves, providing satisfactorily small error bars on the scaled values.  Similarly, memory limitations force us to treat the indium 4d electrons as core, even though they are comparatively 
shallow: about 14.5 eV below the VBE. This leads\cite{NeutralIA} to errors of up to $\sim0.5$ eV, but these are essentially supercell size independent. They can easily be estimated in, say, the 64 atom cell and added back onto the scaled $E_{\rmd}^{\infty}$ at the end.
Our optimized LDA lattice constant using these chosen pseudopotentials\cite{ZVP} is 
5.827 \AA~and the band gap is 0.667 eV, compared to 5.869 \AA~ and 1.344 
eV in experiment. We use $\mu_{\hbox{\scriptsize P}}$ = 
3.485 eV and $\mu_{\hbox{\scriptsize In}}$ = 6.243 eV,
corresponding to stoiciometric conditions, together with
$\mu_{\hbox{\scriptsize Zn}}$ = 1.891 eV,
$\mu_{\hbox{\scriptsize Si}}$ = 5.977 eV and
$\mu_{\hbox{\scriptsize S}}$ = 4.600 eV.
For the 64 atom cell a planewave cutoff energy of 200 eV and a Monkhorst-Pack 4x4x4 k-point grid 
\cite{Monkhorst-Pack} was previously found \cite{ZVP} sufficient to restrict errors to $\Or$(0.01 eV) or less.  When analyzing the errors arising from the supercell approximation itself, non-finite size dependent errors\cite{NeutralIA} (from the In pseudopotential, planewave cutoff etc) are not a problem. However, we do need to keep the k-point sampling errors down to at least the meV scale, since this convergence rate varies with supercell size. This is a much higher convergence criterion than is normally practical, necessary or even meaningful, and it is the reason that we pick only a limited number of example defects for this study.
This convergence level can be achieved\cite{NeutralIA} by using the average 
\begin{equation}
{\overline{E_{\rmd}^{\rmC}}} = \frac{\sum_{N} N^3 E_{d}^{C}\!\left(N\right)}{\sum_{N} N^3}
\end{equation}
\noindent weighted by the number of points in the full Brillouin zone, where $E_{d}^{C}\!\left(N\right)$ is the formation energy calculated using an $N\times N\times N$ Monkhorst-Pack k-point grid. The sum over $N$ is taken up to 12 in the 8 atom cell, 8 in the 64 atom cell and 4, or for certain cases 6, in the 216 and 512 atom supercells in the unrelaxed geometries. (The weighted mean ${\overline{E_{\rmd}^{\rmC}}}$ converges much faster than the unweighted mean or the individual values $E_{d}^{C}\!\left(N\right)$ themselves.) 

We present both non-relaxed (ions at ideal lattice sites) and relaxed calculations. No restrictions are placed upon the symmetry of relaxations, but we do not allow atoms located on the surface of the 
cell to relax.  The relaxation energy
\begin{equation}
\epsilon_{\hbox{\scriptsize Relax}}\!\left(N\right) = 
E_{\rmd}^{\rmC\rmRx}\!\left(N\right) - 
E_{\rmd}^{\rmC\rmId}\!\left(N\right)
\end{equation}
\noindent (where $E_{\rmd}^{\rmC\rmRx}\!\left(N\right)$ and 
$E_{\rmd}^{\rmC\rmId}\!\left(N\right)$ are 
$E_{\rmd}^{\rmC}\!\left(N\right)$ with atoms at relaxed and ideal 
positions respectively) converges faster with $N$ than either 
$E_{\rmd}^{\rmC\rmRx}\!\left(N\right)$ or
$E_{\rmd}^{\rmC\rmId}\!\left(N\right)$. Hence we save computational time by approximating the  relaxed formation energies 
$E_{\rmd}^{\rmC\rmRx}$ by
\begin{equation}
{\overline{E_{\rmd}^{\rmC\rmRx}}} \approx 
{\overline{E_{\rmd}^{\rmC\rmId}}} -
\epsilon_{\hbox{\scriptsize Relax}}\!\left(N\right)
 			   = {\overline{E_{\rmd}^{\rmC\rmId}}} + 
E_{\rmd}^{C\rmRx}\!\left(N\right) - E_{\rmd}^{\rmC\rmId}\!\left(N\right)
\end{equation}
\noindent The relaxation energies used are weighted averages using 
6x6x6 and 8x8x8 k-point grids in the 8 atom cell, 2x2x2 
and (if the convergence is uncertain) 4x4x4 grids in the 64 atom cell 
and 2x2x2 in the 216 and 512 atom supercells.  For the latter cells we 
usually restrict the k-point grid to the irreducible Brillouin zone of 
the undisturbed bulk lattice.  In other words, we use just the special 
k-point (0.25,0.25,0.25): the first Chadi-Cohen k-point\cite{Chadi-Cohen}. This restriction is equivalent to assuming that the distortion 
in the bandstructure due to the presence of the defect is either 
localized (thus important only very near $\Gamma$) or symmetric.  It 
introduces a small error whose significance again disappears in the large 
supercell limit.

\begin{figure*}
\epsfxsize 18.0truecm
\epsfbox[2 9 839 590]{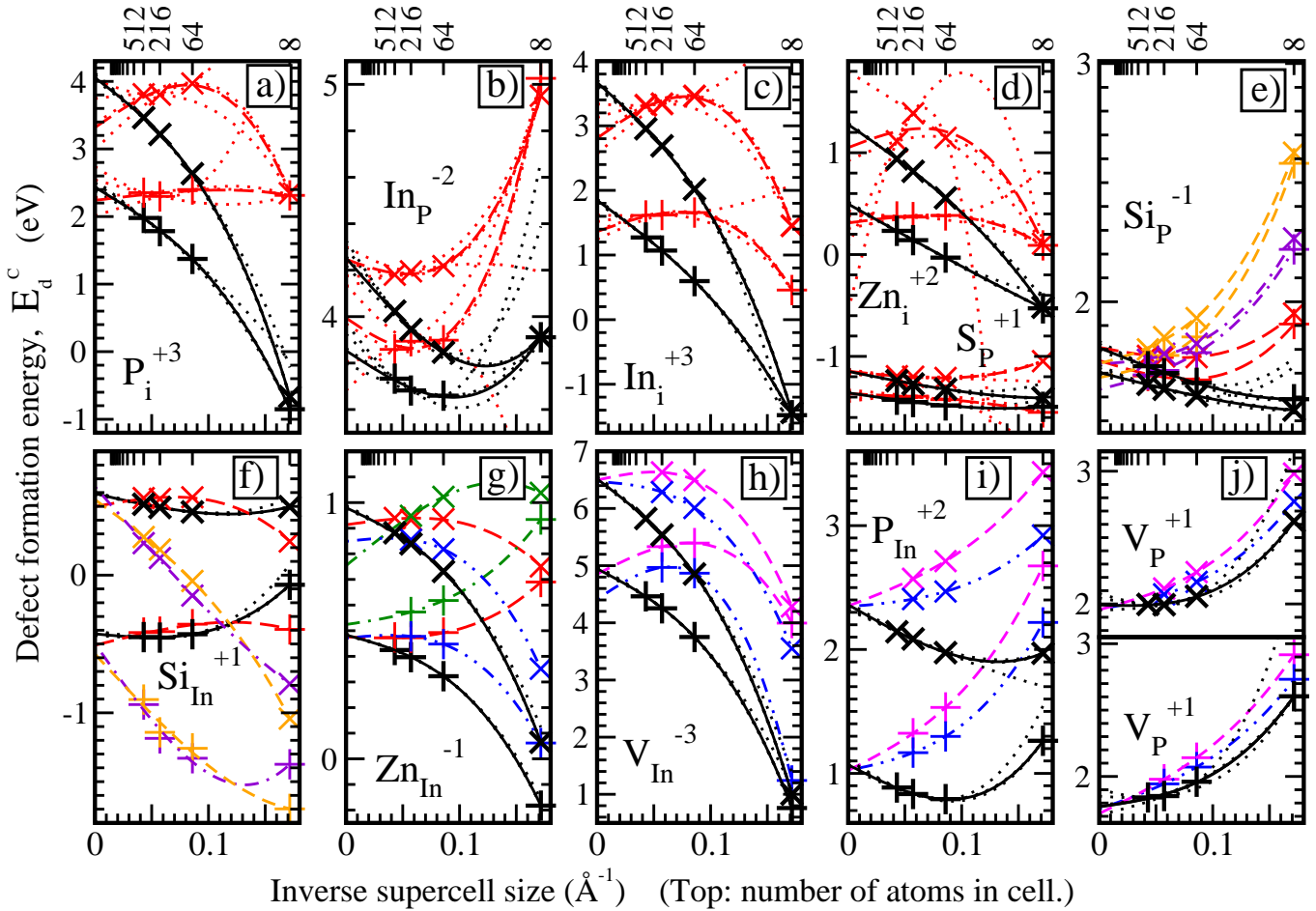} 
\caption{\label{scaling} (Colour online.) Scaling of ($\times$) unrelaxed and ($+$) relaxed formation energies. Curves are fits to Eq \ref{Scale_eqn} with $n$=3. Solid (black) curves are fits to the four points as calculated (no corrections.) Dotted (black) lines each have one cell omitted for accuracy assessment. Scaling of the calculated values with various correction factors are shown for certain examples, as follows. 
Potential re-alignment: long dashed (red) lines in panels a) to g), and accuracy assessment for them: dotted (red) lines in panels a) to d). 
Dispersion corrections: short dashed (orange) lines in e) and f). 
Dispersion+potential corrections combined: dot-dashed (purple) lines in e) and f).
First order ($L^{-1}$) Makov-Payne corrections as dot-dot-dashed (blue) lines in panels g) to j). 
First+third order ($L^{-1}$ + $L^{-3}$) Makov-Payne corrections: short dashed (pink) lines in h) to j).
First order Makov-Payne+potential corrections combined: dash-dash-dot (green) lines in panel g).}
\end{figure*}

\section{\label{Scaling section} Finite Size Scaling of Defect Formation Energies.}

\subsection{Scaled formation energies for the example defects.}

Fig \ref{scaling} shows the formation energy scaling for the 11 example defects in InP. The scaling curves using the uncorrected, as-calculated values are shown as solid lines in the figures (black in the online colour version). Their y-axis intersects give the $E_{\rmd}^{\infty}$ values listed in table \ref{Scaled E}. The curves also serve to predict the formation energy which would be expected in any finite sized supercell: for example the formation energies in the 8000 atom supercell are those at 1/L = 0.1 in the figures.
We can estimate how accurate the $E_{\rmd}^{\infty}$ values are by adding the four dotted (black) curves shown for each example in Fig \ref{scaling}, in each of which one of the four data points has been omitted. (Note that for some cases the errors are so small that the dotted lines are hard to pick out, but they are still present in the figure.) The spread in y-axis intersects gives the error bars listed in the table. This is one of the particular advantages of using finite size scaling: it is possible not only to correct for the finite size errors themselves, but also to obtain a well defined uncertainty on the resulting energies - something other correction schemes can not provide. 

The errors obtained are on the 0.01-0.1 eV range or below, (smaller error are here rounded to 0.01 eV) and can doubtless be further improved if still large supercells are used. Note that, by construction, the errors which arise if only the 8, 64, and 
216 atom supercells are used for the scaling are also on this 0.01-0.1 eV level (See table \ref{Scaled E}.)

The fact that such small error bars can be obtained indicates that a) scaling is a viable and practical approach to supercell approximation errors, b) the k-point convergence is sufficient for our current purpose and c) our enforced use of the 8 atom supercell is actually reasonable, for the same reasons described above and previously\cite{NeutralIA}.

\begin{table*}
\caption{\label{Scaled E}Scaled relaxed and unrelaxed (ideal 
lattice sites) formation energies with ($E_{\rmd}^{\infty,\phi}$) and without ($E_{\rmd}^{\infty}$) potential corrections, for various example defects in InP.  
Note that the error bars are not actually symmetric: the widest has been listed in each case. 
$\epsilon\!(L^{-1})$ and $\epsilon\!(L^{-3})$ are ab initio values of the dielectric constant $\epsilon$ for InP, calculated by comparing the Makov-Payne corrections of order $L^{-1}$ and $L^{-3}$ with the coefficients $a_1$ and $a_3$ obtained from the scaling. $\epsilon^{\phi}\!(L^{-1})$ is the same thing calculated from the potential-corrected formation energies. All energies in eV, dielectric constants in units of the free space dielectric constant $\epsilon_0$.}
\begin{center}
\begin{tabular}{lrrrrrrrrrrrrrrrr}
\hline\hline
 		&\multicolumn{7}{c}{Ideal structures}              				&&\multicolumn{7}{c}{Relaxed structures}					\\\hline
Defect
&\multicolumn{1}{c}{$E_{\rmd}^{\infty}$}	
&\quad
&\multicolumn{1}{c}{$E_{\rmd}^{\infty,\phi}$}
&\quad
&$\epsilon\!(L^{-1})$		
&$\epsilon\!(L^{-3})$
&$\epsilon^{\phi}\!(L^{-1})$
&$\qquad$
&\multicolumn{1}{c}{$E_{\rmd}^{\infty}$}
&\quad
&\multicolumn{1}{c}{$E_{\rmd}^{\infty,\phi}$}
&\quad
&$\epsilon\!(L^{-1})$		
&$\epsilon\!(L^{-3})$
&$\epsilon^{\phi}\!(L^{-1})$\\\hline 
\VPq   	& 1.95$\pm$0.09	&&2.03$\pm$0.01	&&77.21	&8.37	&64.17	&& 1.77$\pm$0.14	&&1.81$\pm$0.03	&&-16.38	&4.32	&-17.41	\\
\VIq   	& 6.52$\pm$0.06  	&&5.75$\pm$0.78	&&12.12  	&-12.86	&-14.43	&& 4.95$\pm$0.05	&&4.63$\pm$0.52	&&17.96	&-16.91	&-39.93	\\
\PIq    	& 2.36$\pm$0.04      	&&2.04$\pm$0.31	&&15.63	&37.37	&-14.88	&& 1.07$\pm$0.04	&&0.83$\pm$0.27	&&19.83	&17.96	&-18.21	\\
\IPq    	& 4.25$\pm$0.08      	&&4.25$\pm$0.12	&&14.33	&29.16	&46.67	&& 3.85$\pm$0.13	&&4.00$\pm$0.31	&&25.53	&30.12	&37.69	\\
\PiPq   	& 4.05$\pm$0.07      	&&3.32$\pm$0.71	&&14.22   	&15.49	&-15.83	&& 2.43$\pm$0.11	&&2.24$\pm$0.50	&&18.08	&-26.71	&-90.00	\\
\IiPq   	& 3.67$\pm$0.08      	&&2.80$\pm$0.54	&&8.18     	&-15.78	&-14.79	&& 1.85$\pm$0.04	&&1.36$\pm$0.25	&&14.18	&215.60	&-29.86	\\
\ZniPq   	& 1.28$\pm$0.01      	&&1.05$\pm$0.31	&&10.57	&-22.52	&-18.98	&& 0.50$\pm$0.02	&&0.31$\pm$0.12	&&13.58	&472.79	&-54.67	\\
\ZnIq   	& 0.98$\pm$0.01      	&&0.91$\pm$0.07	&&9.78	&-13.23	&-32.65	&& 0.48$\pm$0.01	&&0.47$\pm$0.03	&&16.59	&-9.36	&117.20	\\
\SiPq   	& 1.82$\pm$0.03      	&&1.85$\pm$0.03	&&10.85	&42.10	&22.63	&& 1.71$\pm$0.04	&&1.75$\pm$0.13	&&16.65	&87.59	&8.78	\\
\SPq   	&-1.17$\pm$0.02     	&&-1.15$\pm$0.03	&&9.24	&5.38	&16.91	&&-1.34$\pm$0.01	&&-1.37$\pm$0.07	&&12.28	&6.71	&39.50	\\
\SiIq   	& 0.62$\pm$0.01      	&&0.50$\pm$0.11	&&10.81	&20.48	&-13.90	&&-0.36$\pm$0.03   &&-0.51$\pm$0.11	&&27.56	&8.89	&-6.82	\\\hline
Average	&         $\pm$0.05	&&        $\pm$0.27	&&17.54	&8.54	&2.27	&&         $\pm$0.05	&&        $\pm$0.21	&&15.08	&71.91	&-4.88	\\\hline
\multicolumn{9}{r}{\bf Average over both relaxed and unrelaxed structures:}		 &    $\bf\pm${\bf0.05}    &&$\bf\pm${\bf0.24}&&{\bf16.31}	&{\bf40.23}&{\bf-1.31}\\\hline\hline
\end{tabular} 
\end{center}
\end{table*}

\begin{table*}
\caption{\label{E errors}Assessment of correction schemes: Finite size errors (relative to the scaled values) are shown for the 64 atom supercell:  $\delta_{\hbox{\tiny{\it E}}}$ is the error in the as-calculated formation energy, $\delta_{\hbox{\tiny{\it E}+1}}$ and $\delta_{\hbox{\tiny{\it E}+1+3}}$ are the errors when Makov-Payne corrections are used to order $L^{-1}$ and $L^{-3}$ respectively. $\delta_{\hbox{\tiny{\it E}+1}}^{\hbox{\tiny{LDA}}}$ is the error when order $L^{-1}$ corrections are used, calculated with the ab initio dielectric constant evaluated from the results themselves (see text.) $\delta_{\hbox{\tiny{\it E}+k}}$ is the error when defect level dispersion is corrected for in the ionized states of the shallow donors and acceptors. Columns $\delta_{\hbox{\tiny{\it E}}}^{\phi}$ etc are the same as $\delta_{\hbox{\tiny{\it E}}}$ etc but electrostatic potential realignments added. The averages are of the absolute error values $\left|\delta_{\hbox{\tiny{\it E}}}\right|$. All energies in eV.}
\begin{center}
\begin{tabular}{lrrrrrrrrrrrrrrrrrrrrrrrrrr}
\hline\hline
 		&\multicolumn{12}{c}{Ideal structures}              				&&\multicolumn{12}{c}{Relaxed structures}					\\\hline
Defect
&$\delta_{\hbox{\tiny{\it E}}}$	
&$\quad$
&$\delta_{\hbox{\tiny{\it E}+1}}$	
&$\delta_{\hbox{\tiny{\it E}+1+3}}$
&$\delta_{\hbox{\tiny{\it E}+1}}^{\hbox{\tiny{LDA}}}$
&$\quad$
&$\delta_{\hbox{\tiny{\it E}}}^{\phi}$
&
&$\delta_{\hbox{\tiny{\it E}+1}}^{\phi}$	
&$\quad$
&$\delta_{\hbox{\tiny{\it E+k}}}$	
&$\delta_{\hbox{\tiny{\it E+k}}}^{\phi}$	
&$\qquad$
&$\delta_{\hbox{\tiny{\it E}}}$	
&$\quad$
&$\delta_{\hbox{\tiny{\it E}+1}}$	
&$\delta_{\hbox{\tiny{\it E}+1+3}}$
&$\delta_{\hbox{\tiny{\it E}+1}}^{\hbox{\tiny{LDA}}}$
&$\quad$
&$\delta_{\hbox{\tiny{\it E}}}^{\phi}$
&
&$\delta_{\hbox{\tiny{\it E}+1}}^{\phi}$	
&$\quad$
&$\delta_{\hbox{\tiny{\it E+k}}}$	
&$\delta_{\hbox{\tiny{\it E+k}}}^{\phi}$\\\hline 
\VPq   	& 0.11	&&0.29	&0.18	&0.22	&&0.02	&&0.20	&&		&		&&0.19	&&0.37	&0.30	&0.34	&&0.11	&&0.29	&&		&		\\
\VIq   	&-1.67	&&-0.02	&-0.51	&-0.31	&&-0.13	&&1.51	&&		&		&&-1.20	&&0.44	&-0.08	&0.16	&&-0.01	&&1.63	&&		&		\\
\PIq    	& -0.39	&&0.35	&0.11	&0.23	&&-0.02	&&0.72	&&		&		&&-0.28	&&0.46	&0.23	&0.34	&&0.04	&&0.78	&&		&		\\
\IPq    	&-0.40	&&0.33	&0.08	&0.21	&&-0.03	&&0.70	&&		&		&&-0.19	&&0.54	&0.30	&0.42	&&0.05	&&1.16	&&		&		\\
\PiPq   	&-1.41	&&0.23	&-0.28	&-0.05	&&-0.08	&&1.56	&&		&		&&-1.06	&&0.58	&0.05	&0.30	&&-0.03	&&1.61	&&		&		\\
\IiPq   	&-1.65	&&-0.01	&-0.50	&-0.29	&&-0.21	&&1.44	&&		&		&&-1.25	&&0.39	&3.48	&0.11	&&-0.19	&&1.45	&&		&		\\
\ZniPq   	&-0.72	&&0.01	&-0.16	&-0.11	&&-0.13	&&0.60	&&		&		&&-0.53	&&0.21	&-0.03	&0.08	&&-0.12	&&0.62	&&		&		\\
\ZnIq   	&-0.25	&&-0.07	&-0.16	&-0.10	&&-0.04	&&0.14	&&0.19	&0.40	&&-0.17	&&0.02	&-0.03	&-0.01	&&0.01	&&0.19	&&0.05	&0.21	\\
\SiPq   	&-0.16	&&0.02	&-0.03	&0.00	&&-0.04	&&0.14	&&0.16	&0.27	&&-0.10	&&0.08	&0.03	&0.06	&&-0.03	&&0.09	&&0.09	&0.5		\\
\SPq   	&-0.15	&&0.03	&0.00	&-0.02	&&-0.06	&&0.12	&&-0.89	&-0.78	&&-0.14	&&0.04	&0.05	&0.03	&&-0.07	&&0.12	&&-1.07	&-1.02	\\
\SiIq   	&-0.16	&&0.02	&-0.04	&0.02	&&-0.02	&&0.16	&&-0.61	&-0.50	&&-0.07	&&0.11	&0.06	&0.15	&&0.07	&&0.26	&&-0.90	&-0.83	\\\hline
Average	&0.64	&&0.13	&0.19	&0.14	&&0.07	&&0.66	&&0.46	&0.49	&&0.47	&&0.29	&0.42	&0.18	&&0.07	&&0.75	&&0.53	&0.64	\\\hline
\multicolumn{14}{r}{\bf Average over both relaxed and unrelaxed structures:}	   &{\bf0.56}&&{\bf0.21}&{\bf0.31}&{\bf0.16}&&{\bf0.07}&&{\bf0.71}&&{\bf0.50}&{\bf0.57} \\\hline\hline
\end{tabular} 
\end{center}
\end{table*}

\subsection{Form of the scaling.}
\begin{figure}
\epsfxsize 8.5truecm
\epsfbox[9 12 833 539]{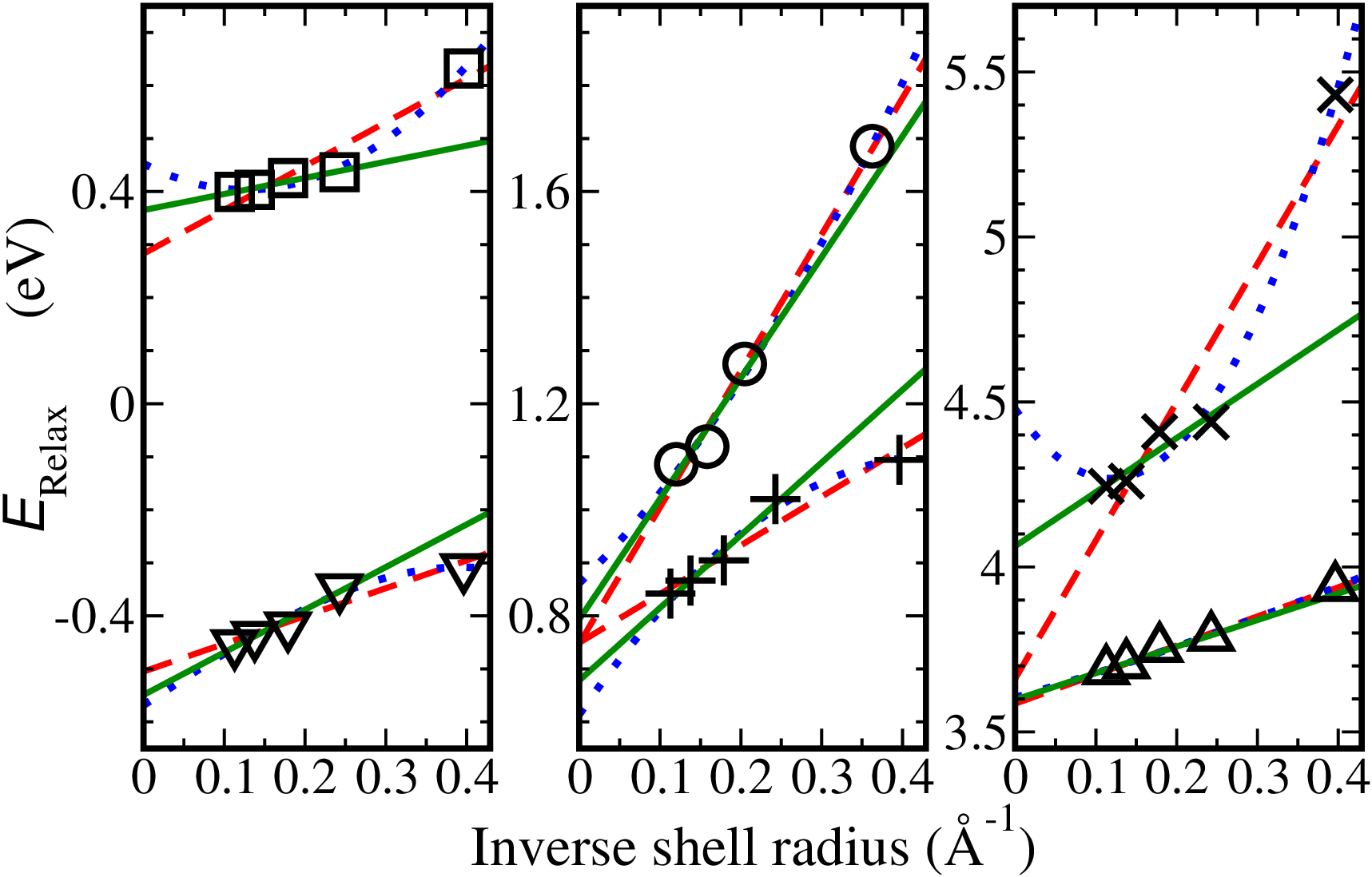} \caption{\label{Shells} (Colour online.) Scaling of the elastic contribution to the finite size errors in defect formation energies. Formation energies in the 216 atom shell are plotted versus the inverse of the radius of the outermost shell of atoms permitted to relax. \ZnIqs ($\Box$), \SiIqs ($\bigtriangledown$), \IiPqs ($\bigcirc$),  \PIqs ($+$), \VIqs ($\times$) and \IPqs ($\bigtriangleup$). Solid (green) and dashed (red) lines: linear fits with the 1 shell only point omitted and included respectively. Dotted (blue) lines: quadratic fits to all points.}
\end{figure}

The choice of $n=3$ again provides the best overall fit to the data, both for relaxed and non-relaxed calculations. Normalized $\chi^2$ tests\cite{NeutralIA} show that on average $n=2$ provides fits 2.9 times worse than $n=3$ whilst $n=4$ is 2.2 times worse. We note, however, that there are additional small (probably $\Or(0.1)$ eV or less) short-ranged errors present which decay exponentially with supercell size. These arise chiefly from the direct overlap of bound defect states with their PBC images and the resulting dispersion of the defect levels.  In the case of relaxed energies some additional short ranged errors can appear because defects in the 8 atom cell are only surrounded by 1 shell of relaxable atoms. The effect of this upon the form of the scaling can be seen in Fig \ref{Shells}. Here we show the scaling of the elastic contribution to the finite size errors. This is done by
calculating formation energies in the 216 atom cell only, so that the electrostatic errors are essentially constant. The number of shells of atoms permitted to relax around the defect is varied and the resulting formation energies are plotted against the inverse of the radius of the outermost relaxed atom shell. Hence the y-intersect corresponds to the formation energy expected if an infinite number of shells are relaxed around the defect, but with the electrostatic errors inherent for the 216 atom supercell. As expected, and as for the neutral defects\cite{NeutralIA}, the elastic errors are predominantly linear. Indeed, if the ``one shell only" point from each curve is omitted then a linear fit 
works perfectly. (Solid lines in Fig \ref{Shells}.) The one shell only point corresponds to relaxations in the 8 atom cell, so we expect that the elastic contribution to the supercell approximation errors scales linearly with supercell size apart from some additional short range errors essentially only affecting the 8 atom cell.

These various short ranged errors have never-the-less only a very limited impact upon the final results, introducing only some additional scatter in the curves in Fig \ref{scaling}, and hence leading to larger error bars in some cases. They also lead to $n=2$ or $n=4$ actually providing the best fit for some individual defects. However, in these latter cases the fitting with $n=3$ is almost always a very close second. These problems can be overcome in a few years time once improved computing 
resources allow the study to be repeated using the 1000 atom supercell. For now we can still 
conclude that the elastic errors are essentially inverse-linear in supercell size, while the total 
formation energy errors (relaxed or unrelaxed) do indeed scale with the inverse-linear dimension and the inverse volume of the supercell.

\section{\label{Corrections section}Assessment of correction schemes.}

In addition to the as-calculated formation energies, Fig \ref{scaling} also shows the formation energy scaling using various correction schemes. For clarity and space we do not show all possible corrections for all example defects, but results for all schemes are listed in tables \ref{Scaled E} and \ref{E errors}. All schemes recover the correct formation energy in the infinite supercell limit, but not all produce improvements over the uncorrected formation energies for smaller supercells. This is shown in table \ref{E errors}, which lists the residual errors (relative to the infinite supercell limit) when the corrections are applied in the 64 atom cell. The uncorrected 64 atom cell formation energies contain average errors of about 0.5-0.6 eV, while using the potential re-alignment scheme produces errors of around 0.1 eV. Makov-Payne does much worse (average errors around 0.1-0.4 eV, but often much larger) and the dispersion ``corrections" produce errors which can be even larger than those in the uncorrected formation energies. 

\subsection{\label{potcorr}Potential Realignment}
The potential re-alignment scheme is illustrated by the long-dashed (red) curves and points in Figs \ref{scaling} a) to g). Even by the 64 atom supercell the values are very good indeed. However, a lot of additional scatter is introduced into the corrected formation energies, $E_{\rmd}^{\rmC,\phi}$. Indeed, the average errors relative to (the uncorrected) $E_{\rmd}^{\infty}$ do not shrink at all with increasing supercell size: 0.07 eV in the 64 atom cell, 0.10 eV in the 216 atom cell and 0.09 eV in the 512 atom cell (see table \ref{E errors}). This leads to wide error bars if the $E_{\rmd}^{\rmC,\phi}$ values are scaled to give the infinite supercell limit, $E_{\rmd}^{\infty,\phi}$. We have derived scaling error bars by the same technique described above. The resulting $E_{\rmd}^{\infty,\phi}$ values and error bars are listed in Table \ref{Scaled E}, although the (red) dotted curves with data points are only shown in Figs \ref{scaling} a) to d). We find error bars of up to $\pm$0.78 eV, average $\pm$0.24 eV. This means that potential realignment is a useful correction for the results from individual supercells, but should {\it not} be used if more accurate results or defined error bars on results are required. In that case non-realigned values should be scaled. These error bars are certainly too large to provide a basis for analysis of other correction schemes. The reason is that the correction scheme, good though it is, is not actually complete or correct. Even in the largest supercells, the point furthest from the defect is {\it not} bulk-like, as the scheme assumes, resulting in either an over estimate or an under estimate, depending upon the specific conditions. 

\subsection{Dispersion Corrections}
The dispersion correction scheme is illustrated for shallow donors and acceptors in Fig \ref{scaling}Êe) and f) and in Table \ref{Scaled E}, both with (short dashed, orange curves) and without (dot-dashed purple curves) potential alignment. Although the acceptor states fare better than the donors, the ``corrected" values are always worse than those using only potential re-alignment, and usually worse than even the uncorrected formation energies. Clearly, even for shallow defect levels, which follow\cite{VdeW+N} closely the VBE or conduction band edge (CBE), $\epsilon_D^{\Gamma}$ still produces worse formation energies than $\epsilon_D^{ks}$. 

\subsection{Makov-Payne Corrections}
Fig \ref{scaling} g) shows the first order $L^{-1}$ Makov-Payne corrections, with (dash-dash-dot, green) and without (dot-dot-dash, blue) potential re-alignment. When used together the two schemes usually produce a large over-estimate of the required correction, (see columns 7 and 15 of table \ref{E errors},) almost as if using both corrections actually makes the {\it same} correction twice. Since the combination does so much worse than either technique alone there is no point going further with it.
Instead, Figs \ref{scaling} h) to j) show Makov-Payne corrections only, with formation energies including both the order $L^{-1}$ corrections (short dashed, magenta) and the order $L^{-1}$ plus order $L^{-3}$ corrections (dot-dot-dashed, blue).
The order $L^{-1}$ corrections work well in some cases (such as \IiPqs when relaxations are omitted), but in most cases they are too large by a factor of about 1$\frac{1}{2}$ to 2, (as also noted by others\cite{MP-fails,MP-fails-OtherKorr}) so that the ``corrected" formation energies are little better than the uncorrected ones. When the order $L^{-3}$ corrections are added the correct formation energies are obtained in some cases, such as \VIq, but in other cases, such as \PIq, they help but are not sufficient. For other cases, such as  \VPq, the corrections actually move the formation energies in the wrong direction.

Table \ref{E errors} shows that the corrections are generally more likely to succeed for unrelaxed formation energies which is to be expected since the electrostatic monopole terms are not the only ones to scale as  $L^{-1}$: the elastic errors do too. This means that even in principle the Makov-Payne corrections are only useful for non-relaxed formation energies, which are rarely the interesting ones. Besides this, the corrections also do better for more highly charged defects. This confirms that one of the problems is that they do not take into account the various other error terms which depend upon supercell size but not on charge state. These errors mostly have to do with the spurious defect level dispersion introduced by the PBCs. Although the direct contributions of these are exponentially decaying\cite{NeutralIA}, their effects can still be seen in supercells on the scale of 10-100 atoms. Indeed the actual band width can be on the order of, for example, 0.5 eV and 2 eV in the 64 atom and 8 atom supercells \cite{AndreasGaP} and remain significant even beyond that.  
Indirect dispersion effects can also be very important: for example, in a partially filled, erroneously dispersed defect level only the lower part will be filled, leading to too low a value for $E_{\rmd}^{\rmC}$. Worse happens if the defect level lies outside the band gap, either because it genuinely does or because the supercell is too small. This can lead to strong linear terms in the supercell size errors even for neutral defects\cite{NeutralIA}: a neutral defect can behave as, say a -1 charged defect with (to a first approximation) a +1 charged jellium background. This is not limited to neutral defects: a calculation for a defect anticipated to be in a +2 charge state (with a -2 charged jellium) could end up behaving more like a +3 charge defect with a -3 charged jellium. If the defect level moves outside the bandgap at certain k-points only it can lead to a linear error term which is not even proportional to the square of an integer charge.
Overall, even the leading linear error term may be very different from that predicted by Makov and Payne's corrections. 
Unfortunately, beyond noting that things get better on average for larger charge states and for non-relaxed calculations 
there seems to be no {\it a priori} method for determining whether the corrections will make things better or worse in a specific case. They are thus of little practical help, since they do not take into account enough of the specific behaviour of individual defects and materials. Indeed, it seems unlikely that any such highly generalized model for prediction of finite size error correction factors will ever fully succeed.

\subsection{\label{Ab initio epsilon section}Calculating the ab initio dielectric constant.} 
Makov and Payne predicted that the two leading terms in the errors should be linear and cubic and our results show that they were correct in that respect. Their ``corrected" formation energy takes the form
\begin{equation}\label{MP_eqn}
E_{\rmdMP}^{\rmC}\!(L)  =  E_{\rmdMP}^{\infty} - k_{1}(\epsilon L)^{-1} - k_{3}(\epsilon L)^{-3}
\end{equation}
\noindent where $\epsilon$ is the dielectric constant, $k_1=q^2\alpha/2$ and $k_3=2\pi qQ/3$. ($q$ is the charge of the defect, $\alpha$ is the Madelung constant for the supercell and $Q$ is the quadrupole moment of the defect.) Comparing this with Eq \ref{Scale_eqn} we find $a_n=-k_n/\epsilon$. If we assume that the scheme is correct after all, then, $q$ being known and $Q$ having been calculated from the charge density, the only variable is the dielectric constant $\epsilon$. We can then use the correction scheme together with the scaling results to derive an {\it ab initio} value of $\epsilon$. This can be done twice for each defect, as shown in table \ref{Scaled E}. We find a wide scatter in the results, reflecting the wide variations in the effectiveness of the corrections. Indeed, the values of $\epsilon$ obtained are completely crazy when order $L^{-3} $Êcorrections are used, as these are much more sensitive to short range effects and other errors. This, again reflects the fact that the situation described by Makov and Payne was highly idealized and ignores too many of the details of the charge distribution around specific defects.
Never-the-less, the averaged values $\epsilon$ from the order $L^{-1}$ corrections are reasonably good. The most physically correct approach is to use the unrelaxed formation energies only, (with no elastic effects); indeed the values derived using relaxed values, third order corrections or Makov-Payne plus potential re-alignment make little sense  (see \ref{Scaled E}). From the first order non-relaxed curves we obtain a dielectric constant of 17.5$\pm$19.0. (The error bar is the standard deviation from the average.) Numerical problems with the \VPs value have made it rather unreliable, and very different to the the others. Omitting it gives the perhaps more consistent value of 11.6$\pm$4.1. These values compare to 9.6 in experiment
or 10.7 calculated\cite{LDA e-constant} using more traditional ab initio DFT-LDA techniques.\cite{LDA e-notes} We thus obtain a fairly reasonable estimate of $\epsilon$ as a free side-effect of performing accurate defect calculations - an interesting alternative to the traditional calculations methods. The uncertainty in the value obtained is obviously rather large, but should improve if more defects in more charge states are included in the average. 

The order $L^{-1}$ Makov-Payne corrections do improve using this new value for the dielectric constant, see columns 5 and 13 of table \ref{Scaled E}. However, some individual values still have errors of up to 0.3-0.4 eV, and there is still no way to know when the corrections are making things better and when they are making things worse, so from a practical point of view the Makov-Payne scheme is still not reliable enough for accurate calculations.

\begin{figure*}
\begin{minipage}[l]{8.0truecm}
\includegraphics[width=8.0truecm]{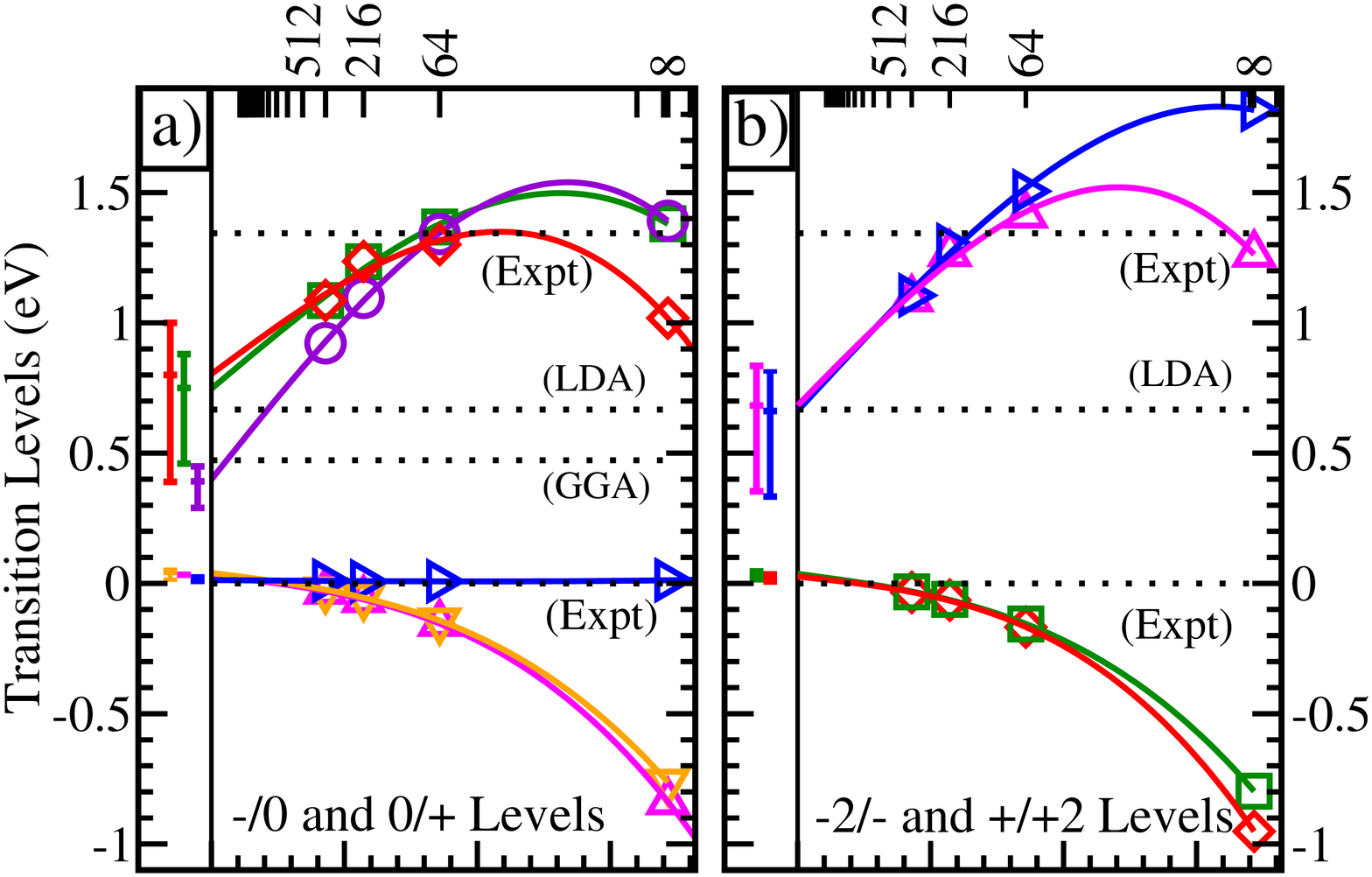}
\vglue 0.5 truecm
\includegraphics[width=8.0truecm]{Figure3cd_clr2.eps}
\end{minipage}
\begin{minipage}[r]{8.0truecm}
\includegraphics[width=8.0truecm]{Figure3ef_clr2.eps}
\vglue 1.3 truecm
\includegraphics[width=8.0truecm]{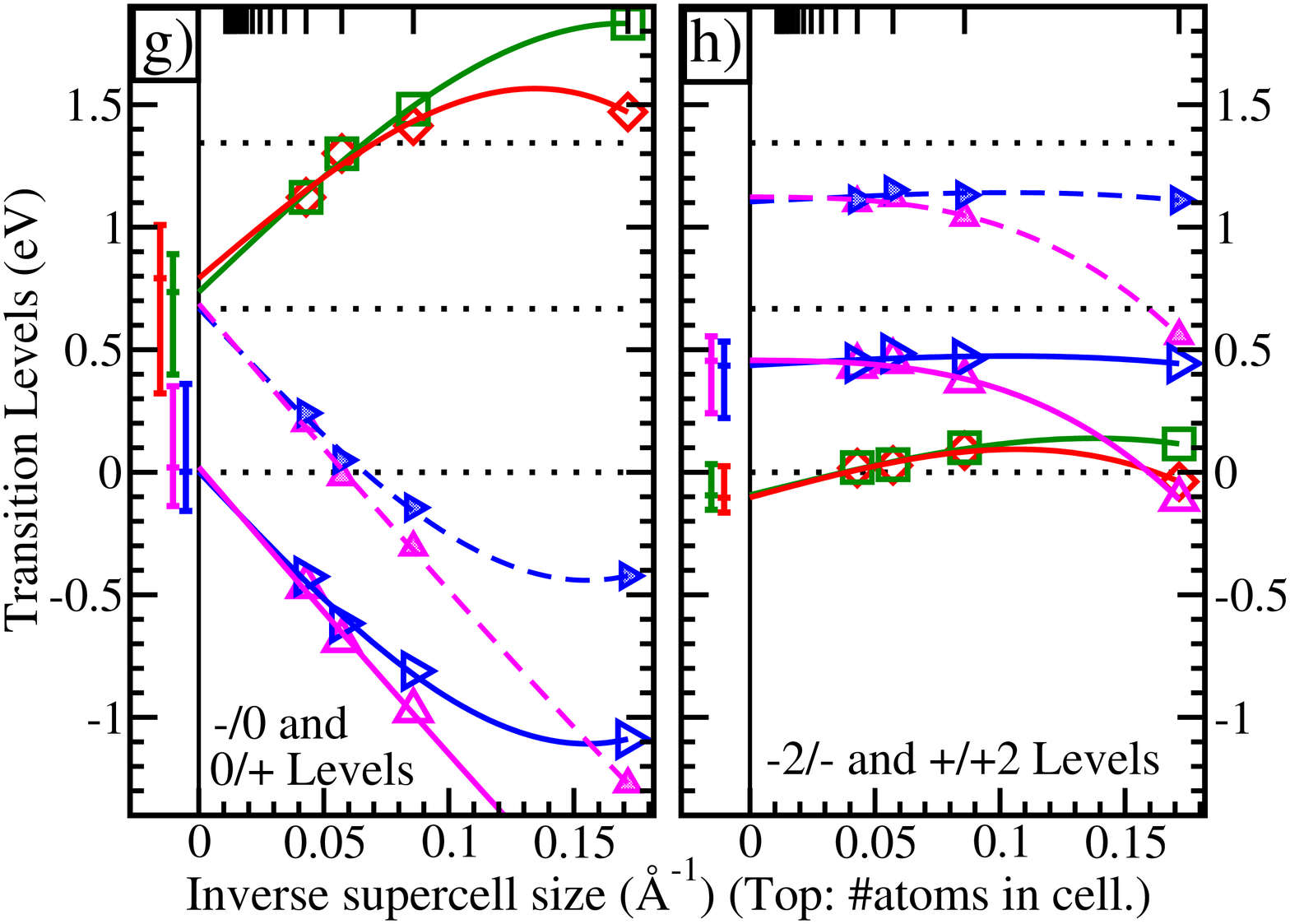}
\end{minipage}
 \caption{\label{Transfer graph} (Colour online.) Scaling of transfer levels for simple donors and acceptors calculated using:  a) \& b): neutral bulk as reference, with no corrections. c) \& d): neutral bulk reference with potential corrections. e) \& f) neutral bulk reference with dispersion corrections. g) \& h) charged bulk as reference, with no corrections. Left panels: the dopant levels themselves. Right panels: the double donor or double acceptor levels, which should lie outside the bandgap. Using LDA: \SPT and \SPDT ($\Box$, green), \SiIT and \SiIDT ($\Diamond$, red), \SiPT and \SiPDT ($\rhd$, blue),  \ZnIT and \ZnIDT ($\bigtriangleup$, pink). Using GGA: \SPT ($\bigcirc$) and \ZnIT ($\bigtriangledown$). In g) \& h) the smaller symbols with dashed lines show the acceptor-type levels relative to the experimental CBE rather than the LDA one. On all panels: the dotted lines are (in order of increasing energy) the VBE and CBE from: GGA (panel a)), LDA and experiment. The error bars shown have been constructed as described in section \ref{Scaling section} though the dotted lines are omitted for clarity.}
\end{figure*}

\section{\label{Bandgap section}The bandgap problem}

We now turn to the band gap problem and the issue of how to map calculated transfer levels onto the experimental gap. In practise several alternative - and essentially incompatible - methods are normally used. 

1) The Extended Gap Scheme: align the theoretical and experimental VBEs and start plotting defect transfer levels from there, continuing {\it past }Êthe theoretical CBE until one reaches the experimental one. In the section of the thus plotted ``band gap" which lies above the theoretical CBE one automatically includes calculations in which supposedly localized, defect-bound electrons are in reality located in delocalized conduction band states. The properties of the defect itself (transfer levels and local relaxed structure etc) re-enter primarily via hybridization of the conduction band states with the localized defect states, though this hybridization becomes smaller as the supercell size grows.

2) The Scissors Scheme: align both the theoretical VBE  {\it and} CBE with their experimental counterparts, performing a ``scissors" operation to stretched out the theoretical gap states over the experimental gap. The manner in which this scissors operation should be done is not uniquely defined. A common option is to place acceptor levels the same distance above the experimental VBE that they appear above the theoretical VBE in calculations, and donor levels the same distance below the experimental CBE that they appear below the theoretical CBE in calculations. A better alternative is to actually examine the form and symmetry of the defect states themselves, and see whether they hybridize more strongly with host states near the CBE or with states near the VBE. If they hybridize most strongly with VBE states then they should be plotted the calculated distance above the (experimental) VBE, and if they hybridize most strongly with CBE states they should be placed the calculated distance below the CBE.

3) The Reference Level Scheme. The basis of this scheme is rather different: the transfer level for the defect of interest is calculated, together with that of a similar reference defect for which the experimental value of the transfer level is well known, both done to the same level of accuracy. The difference between the experimental and calculated levels of the known defect is subtracted from the calculated value of the new defect, so that the new level is only found relative to the old one. This idea is not without practical merit, but is very empirical. Its accuracy depends critically upon the choice of an appropriate reference defect, which must be as similar to the new one as possible, so it will not be discussed further here. However, it has an occasionally used {\it ab initio }Êvariant, which will be discussed:

4) The Charged Bulk Reference Scheme: the reference state is not that of another defect, but is either the VBE or the CBE, meaning that a charged bulk total energy appears in Eq \ref{Eform}, rather than a neutral one. In principle this provides an alternative route around the band gap problem. (Details below.)

Obviously, none of these schemes is fully correct, since the LDA/GGA bandgap problem is a fundamental one, but the important practical question of which approach comes closer to giving the correct physical picture remains unanswered. In principle it can be answered by examining various experimentally well known defect levels. 
The exact location of most native defect levels is rather hard to measure to a sufficiently high accuracy to answer this question, but many simple donor and acceptor levels are known very accurately. We will use the 0/- acceptor level of \ZnI, which in experiment lies 0.035 eV from the VBE, and the +/0 donor levels of \SPs and \SiI, which experiment finds about 0.006 eV from the CBE. We will also add in the 0/- transfer level of \SiP, which would be a simple acceptor if \SiIs had not been the more stable site for Si in InP. This gives us an example of a donor and an acceptor on each sublattice, so that all bonding and band hybridization possibilities are represented.
Unfortunately, calculations of these levels in finite sized supercells in the 100-200 atom range have never produced a clear answer to the question, so we will use finite size scaling to correct for the supercell approximation errors. The results are shown in Fig \ref{Transfer graph} a) using as calculated values, c) adding in potential corrections and e) using dispersion corrections. (Van der Walle and Neugebauer\cite{VdeW+N} suggested that dispersion corrections should still be correct for shallow transfer levels.) The results using as calculated transfer levels and potential corrected ones are very similar. The dispersion corrections, on the other hand, are clearly completely incorrect: they place both acceptor and donor levels in the midgap for most practical supercell sizes, whether the potential corrections are added (not shown) or omitted (as here). Meanwhile, in Fig \ref{Transfer graph} b), d) and f) we also show the 
second donor/acceptor levels, \ZnIDT, \SPDT, etc., calculated using the same correction schemes. Since these levels are never observed experimentally they must lie outside the band gap. Hence the VBE should lie between the double donor levels (right panels) and the single acceptor levels (left panels). Similarly the CBE should lie between the single donor and double acceptor levels. In practise, these pairs of levels more or less coincide, doubtless a result of the remaining limitations in the use of DFT-LDA for semiconductor defects. Fortunately this still leaves us with a clear view of how to treat the band gap problem.

In the 64 atom cell the donor (and double acceptor) levels lie roughly the experimental band gap (1.3 eV) above the VBE, while the acceptor (and double donor) levels lie on average a little below the VBE. However, coming to the larger cells the donor levels fall and the acceptor levels rise. Finite size scaling places the acceptor levels \ZnIT and \SiPT 0.03 and 0.01 eV above the VBE respectively, in rather good agreement with experiment. 
The single donor (and double acceptor) levels all scale to the theoretical CBE.\cite{Note4} To be more specific, transfer levels calculated using LDA scale to the LDA band edges, while the \ZnITs and \SPTs transfer levels calculated using the Perdew-Wang GGA \cite{GGA} scale to the edges of the GGA band gap - Fig \ref{Transfer graph} a). (The GGA CBE lies 0.2 eV below the LDA one when the lattice parameter has been optimized.) 

Hence, scheme 1, the extending gap scheme, is seen to be the most appropriate when only reporting uncorrected results from supercells of about 50-100 atoms. However, when the finite size errors are removed (by scaling or by some other technique) it becomes clear that the scissors scheme, scheme 2, is physically far more correct. Unscaled LDA or GGA results in supercells over a few 1000 atoms would also be best reported using the scissors scheme. For intermediate (100-1000 atom) supercells some kind of hybrid approach is required. The result also indicates why the controversy has lasted so long: ultimately the scissors scheme is correct, but this only shows up for very large supercells or with scaling.\cite{Note1}

We now return to scheme 4), the charged bulk reference. This amounts to replacing the terms  $-E_{\rmT}^{\rmC}\!\left(\hbox{no defect}\right)$ and $-q\epsilon_F$ in Eq \ref{Eform} by the term $-E_{\rmT}^{\rmC}\!\left(\hbox{no defect}^q\right)$, which is the total energy of the bulk supercell ${\rmC}$ with $-q$ extra electrons and neutralizing jellium. Fig \ref{Transfer graph} g) \& h) show the transfer levels calculated like this, with no corrections terms. The donor levels behave in the same way as using Eq \ref{Eform} in Fig \ref{Transfer graph} a), but the acceptor levels are less straight-forward. Using a charged bulk reference the levels come out relative to the CBE, rather than the VBE: they implicitly {\it include} the bandgap, which must be subtracted off again to place them on the same overall scale as the donor levels.  This gives a ``choice" for the value for the bandgap to subtract, which is how the potentially route around the band gap problem enters. Namely, if the 0/- transfer level emerges as, say,  -0.5 eV, one could place it 0.5 eV below the experimental CBE, thus plotting the transfer levels over the experimental bandgap. (Small symbols and dashed scaling curves in Fig \ref{Transfer graph} g \& h).) For the single acceptor levels this clearly does not work: although they land accidentally close to the VBE for smaller supercells they actually scale to the theoretical CBE, which is completely wrong. Instead, they should be placed below the theoretically CBE (large symbols, solid curves), where they scale to the VBE. Unfortunately the opposite is true for the double acceptor levels. These work out roughly OK if plotted relative to the experimental CBE - lying outside the theoretical band gap, even if still inside the experimental one - but using the theoretical CBE, (as required for the single acceptors) they lie inside the theoretical band gap, disagreeing with experiment. Hence using a charged bulk total energy as the reference for charged defect calculations is not even internally consistent and the scheme is thus fundamentally incorrect. 

\section{\label{Conclusions}Conclusions}

In this paper we have shown that finite size errors in the supercell approximation scale with the linear dimension and with the volume of the supercell, and that finite size scaling the results from a series of supercells removes the supercell approximation errors, leaving accurate information on isolated semiconductor defects, without the need for corrections. We also obtain error bars defining the uncertainly on the results obtained, and as far as we are aware this is the only method which is able to remove these errors in a controlled manner with defined uncertainty. We have demonstrated this using a variety of different types of defects with charge states ranging from -3 to +3 and find that it is possible to reduce formation energy errors from the 0.1-2 or so eV range of practical supercells down 
to the 0.01-0.1 eV range or below - doubtless much lower if still larger supercells are used. By construction, errors on this scale also occur if only the 8, 64, and 
216 atom supercells are used.

We then used the scaled results for the first full reliability test of three correction schemes. We found that dispersion corrections are incorrect and Makov-Payne corrections are poor (with both the experimental and LDA dielectric constants,) though they did allow us to obtain a reasonable {\it ab initio} LDA dielectric constant of $\epsilon$ = 11.6$\pm$4.1 for InP. On the other hand, the potential re-alignment scheme was found to be remarkably successful, removing much of the electrostatic defect-defect error as well, to leave average residual errors of about 0.1 eV, from single calculations with supercells in the 64-512 atom range. 

This obviously raises the question of why the potential re-alignment scheme is {\it so }Êsuccessful, when it does not set out to correct defect-image interaction errors at all! The fact that it produce similar (but more reliable) corrections to the Makov-Payne scheme suggests that it is some how dealing with the electrostatic errors anyway. We noted in section \ref{potcorr} that the scheme assumes that the real-space potential at some point in the cell far from the defect is bulk-like, even though for practical cell sizes it is not bulk-like at all. The resulting additional shift in this local real-space potential reflects the effects of the electrostatic defect-image interactions. Doing the potential re-alignment in this way therefore fails to properly correct the mismatch in the zeros of the energy scales between the bulk and defect cells, but the ``error" in the re-alignment more or less corrects for the electrostatic errors arising from the PBCs.

Finally, we have given the long awaited answer to the dilemma of how best to map LDA and GGA calculated defect transfer levels onto the experimental gap, and indicated why the issue was previously so hard to settle. The key result is that the scissors method is physically more correct, though the extended gap scheme is best when reporting results from single supercells on the 1-200 atom scale without finite size corrections. For uncorrected results from supercells over a few 1000 atoms the scissors method is best, with a hybrid method needed in between. The best, of course, is to use the scissors scheme, with either scaled or corrected results, regardless of supercell size. The apparent success of the essentially incorrect extended gap scheme for uncorrected results in manageably sized supercells is the basic reason for the debate lasting so long.

This leads to another issue which is also apparent from our results: It is very dangerous to report calculations from single supercells without trying to estimate the errors contained. Quantitatively these can be $\sim$1-2 eV or more, but we have cases here where conclusions are even qualitatively wrong in supercells up to and even including the 512 atom cell. For example\cite{Note2}, comparing these results for \PIqs with those\cite{NeutralIA} for \PIns we find that even at the CBE, the +2 charge state appears more stable than the +0 in all four supercells. The fact that it is actually 0.19 eV {\it less} stable only emerges when the finite size errors are removed, either by scaling or (leaving residual errors from 0.05-0.13 eV) by using potential realignment. Similarly, at the VBE, \VIqs and \IiPqs are more stable than \VIns and \IiPn, respectively, in both the 64 and 216 atom supercells. The correct stability order only appears in the 512 atom cell, (neutrals more stable by 0.21 and 0.16 eV respectively,) and the correct order of magnitude for the difference (0.68 and 1.17 eV) is only obtained by scaling. Another striking example is that, according to LDA in cells $\le$512 atoms, p-type Zn doped InP - a material upon which much of current optoelectronics depends - should not be p-type at all! For the roughly stoiciometric conditions of, say, Czochralski growth, LDA in the 64 and 216 atom cells places Zn not as the shallow acceptor \ZnIs but as the interstitial \ZniP, where it is a deep double donor. Even in the 512 atom cell the two are degenerate, suggesting at best semi-insulating material. According to this Zn is only an acceptor for InP grown under strongly non-equilibrium conditions, such as with molecular beam epitaxy. However, Zn {\it is }Êa p-type dopant, even grown from the melt, and this fact {\it can }Êbe predicted using LDA, but only for supercells of the order of 1000s of atoms, or if the supercell size errors are removed - by scaling or otherwise. Doing this using potential realignment works for all these examples: even in the 64 atom supercell reasonable results can be obtained. However, caution should still be used:  Firstly, for our examples it worked much better for the formation energies than for the shallow dopant transfer levels. Secondly, potential re-alignment makes \PIns (correctly) more stable than \PIqs at the CBE in all but the 8 atom supercell, but if those corrected results are then scaled the wrong answer returns, with \PIqs more stable than \PIns because of the large error bars found when scaling potential re-aligned energies. 

In short, it is essential, to estimate and report the finite size errors for each specific case when reporting supercell defect calculations. This is often omitted, or is only done using the unreliable Makov-Payne scheme. When it is done this is usually by doing most calculations in a cell of, say, 50-200 atoms, and then repeating a few of them in a slightly larger cell. If the calculated results do not change much then they are considered converged. However, even this should be done with extreme caution. Even with only a linear contribution, the finite size errors in the 64 atom supercell are three times the difference between the 64 and 216 atom cell energies, the 216 atom cell errors are still twice this estimate. Even the errors in the 512 atom cell are three times the difference between the 216 and 512 atom energies. 

So, how {\it should} finite size errors within the supercell approximation be treated? Ideally, using finite size scaling of otherwise {\it un}corrected energies. This is, of course, costly in both human and computer time. The best alternative is simply to use potential realignment in as large a supercell as time and resources permit. However, one should be aware that a) this does not help the elastic errors, b) potential realignment should {\it not }Êbe combined with finite size scaling and c) there is no way to estimate the remaining errors or the reliability of the results. For our examples, the average errors using this method are $\sim$0.10 eV, but with some examples up to 0.21 eV, and nothing to say that much larger errors will never occur. If the conclusions being drawn from a calculation are not adversely affected by uncontrolled errors of 0.1-0.2+ eV then this method is reasonably good. Otherwise, the only truly reliable method of controlling the errors in the supercell approximation, and defining the uncertainly in the results, is finite size scaling. 

The calculations in this paper were performed at Uppsala University and 
at the Parallel Computing Centre (PDC) Stockholm, Sweden.  The authors 
would also like to thank the G\"oran Gustafsson Foundation, the 
Swedish Foundation for Strategic Research (SSF) and the Swedish 
Research Council (VR) for financial support.

$\dagger$ Present address: Materials Chemistry, Box 538, SE-75121 Uppsala, Sweden. Email address: Christopher.Castleton@mkem.uu.se.


\begin{thebibliography}{9}
\bibitem{DFT} W. Kohn and L. Sham Phys. Rev. {\bf 140}, A1133 (1965)
\bibitem{Payne Review}  M.C.Payne, M.P.Teter, D.C.Allan, T.A.Arias and J.D.Joannopoulos Rev. Mod. Phys. {\bf 64}, 1045 (1992) 
\bibitem{Jellium} M.Leslie and M.J.Gillian J. Phys.C: Solid State Phys. {\bf 18}, 973 (1985)
\bibitem{NeutralV} C.W.M. Castleton and S. Mirbt Physica B {\bf 340-342}, 407 (2003)
\bibitem{NeutralIA} C.W.M. Castleton and S. Mirbt Phys. Rev. B {\bf 70}, 195202 (2004)
\bibitem{MP-fails-OtherKorr} P.A. Schultz Phys Rev Lett {\bf 84} 1942 (2000);
	U.Gerstmann, P.De\'ak, R.Rurali, B.Aradi, Th.Frauenheim and H.Overhof Physica B {\bf 340-342}, 190 (2003);
	B.Aradi, P.De\'ak, A.Gali, N.T.Son and E.Janz\'en Phys Rev B {\bf 69}, 233202 (2004)
\bibitem{OtherKorr} L.N. Kantorovich Phys. Rev. B {\bf 60}, 15476 (1999);
         L.N. Kantorovich and I.I.Tupitsyn J. Phys.:Condens. Matter {\bf 11}, 6159 (1999);
	P.A. Schultz Phys Rev B {\bf 60}, 1551 (1999);
         H. Nozaki and S. Itoh Phys. Rev. E {\bf 62}, 1390 (2000);
         A.Castro, A.Rubio and M.J.Stott Can. J. Phys. {\bf 81}, 1151 (2003)
\bibitem{MP} G. Makov and M.C. Payne Phys. Rev. B {\bf 51}, 4014 (1995)
\bibitem{MP-fails} J.Lento, J.-L..Mozos and R.M.Nieminen J. Phys.:Condens. Matter {\bf 14}, 2637 (2002)
\bibitem{VdeW+N} C.G.Van de Walle and J.Neugebauer J. App. Phys. {\bf 95}, 3851 (2004)
\bibitem{Wei} S.-H. Wei Comp. Mat. Sci. {\bf 30}, 337 (2004)
\bibitem{AndreasGaP} A. H\"oglund, C.W.M. Castleton and S. Mirbt Accepted to appear in Phys. Rev. B (2005)
\bibitem{Quasiparticles} M.Hedstr\"om, A.Schindlmayr and M.Scheffler Phys. Stat. Sol. B {\bf 234}, 346 (2002)
\bibitem{Previous InP} R.W.Jansen Phys. Rev. B {\bf 41}, 7666 (1990);
	A.P.Seitsonen, R.Virkkunen, M.J.Puska and R.M.Nieminen Phys. Rev. B {\bf 49}, 5253 (1994);
	A. H\"oglund, M.C.Qian, C.W.M. Castleton, M.G\"othelid, B.Johansson and S. Mirbt In preparation. (2005)
\bibitem{USPP} D. Vanderbilt Phys. Rev. B {\bf 41}, R7892 (1990);
         G. Kresse and J. Hafner J. Phys: Cond. Matt. {\bf 6}, 8245 (1994)
\bibitem{VASP} G. Kresse and J. Furthm\"uller Comp. Mat. Sci. {\bf 6}, 15 (1996)
\bibitem{ZVP} C.W.M. Castleton and S. Mirbt Phys. Rev. B {\bf 68}, 085203 (2003)
\bibitem{Monkhorst-Pack} H. Monkhorst and P. Pack Phys. Rev. B {\bf 13}, 5188 (1976)
\bibitem{Chadi-Cohen} D.J.Chadi and M.L.Cohen Phys. Rev. B {\bf 8}, 5747 (1973)
\bibitem{LDA e-constant} B.Arnaud and M.Alouani Phys. Rev. B {\bf 63}, 085208 (2001)
\bibitem{LDA e-notes} The dielectric constant calculation quoted here used plain LDA with no added local correlation effects, GW quasi-particle corrections or scissors shifts of the bandstructure, and so comes from a similar level of method to ours.
\bibitem{Note4} The scaling of the single donor and double acceptor levels produces rather wide error bars. The reason is that the particular Kohn-Sham level which is half filled in the 0 charge state but empty in the +1 charge state lies above the CBE in all of these (finite) supercells, though it will re-appear inside the gap for large enough cells. As we showed previously\cite{NeutralIA} this tends to produce poor scaling in the 0 charge state. 
\bibitem{GGA} J.P. Perdew and Y. Wang Phys. Rev. B {\bf 13}, 5188 (1976)
\bibitem{Note1} The data shown in figure \ref{Transfer graph} were calculated using the In 4d electrons as core, but the conclusions are completely unaffected: correcting this error using In 4d valence calculations in the 64 atom cell, the curve for \ZnITs moves down by about 0.003 eV, that for \SiPTs moves up about 0.04 eV and those for \SiITs and \SPTs move down about 0.08 eV. We also note here that the present results are for relaxed formation energies, but that we can reach exactly the same conclusions using non-relaxed formation energies.
\bibitem{Note2}The numerical results reported in this paragraph include corrections for the use of the In 4d core pseudopotential. They are thus considered accurate to around 0.01-0.02 eV, not counting errors in the LDA itself.
\end{thebibliography}
\end{document}